\documentclass{caosp309}

\usepackage{graphicx}

\usepackage{natbib}
\articleNo{301}
\pubyear{2020}
\volume{50}
\volnumber{3}
\firstpage{1}
\received{May 1, 2020}
\accepted{July 28, 2020}

\def\BibTeX{{\rm B\kern-.05em{\sc i\kern-.025em b}\kern-.08em
             T\kern-.1667em\lower.7ex\hbox{E}\kern-.125emX}}

\newcommand{\fink}{{\sc Fink}}

\begin{document}

%
\htitle{Satellite glints in ZTF and LSST}
\hauthor{S.\,Karpov and J.\,Peloton}

\title{The rate of satellite glints in ZTF and LSST sky surveys}


%
%
\author{
        S.\,Karpov\inst{1}\orcid{0000-0003-0035-651X}
      \and
        J.\,Peloton\inst{2}\orcid{0000-0002-8560-4449}
       }

%
\institute{
    CEICO, Institute of Physics of Czech Academy of Sciences, Na Slovance 1999/2, 182 00 Prague 8, Czech Republic, \email{karpov@fzu.cz}
    \and
    Universit{\'e} Paris-Saclay, CNRS/IN2P3, IJCLab, Orsay, France
    }

\date{March 8, 2003}

\maketitle

\begin{abstract}
We assess the impact of satellite glints -- rapid flashes produced by reflections of a sunlight from flat surfaces of rotating satellites -- on current and future deep sky surveys such as the ones conducted by the Zwicky Transient Facility (ZTF) and the Vera C. Rubin Observatory upcoming Legacy Survey of Space and Time (LSST). In addition to producing a large number of streaks polluting the images, artificial satellites and space debris also generate great amount of false point-source alerts hindering the search for new rapid astrophysical transients. To investigate the extent of this problem, we perform an analysis of isolated single frame events detected by ZTF in more than three years of its operation, and, using three different methods, assess the fraction of them related to artificial satellites to be at least 20\%. The satellites causing them occupy all kinds of orbits around the Earth, and the duration of flashes produced by their rotation is from a fraction of a second down to milliseconds, with mean all-sky rate of up to 80,000 per hour.

%
\end{abstract}

%
\section{Introduction}
\label{sec:introduction}

Modern large-scale time-domain sky surveys, like upcoming Vera C. Rubin Observatory Legacy Survey of Space and Time \citep{lsst} and ongoing Zwicky Transient Facility \citep{Bellm_2018} offer an unprecedented possibility of not only detailed study of already known classes of transient objects, like variable stars, novae, supernovae, tidal disruption and microlensing events, AGNs and distant Solar System objects, but also of discovery of potentially new and exciting classes of astrophysical transients, especially in the still poorly studied region of shortest time scales. Indeed, despite long history of various experiments aimed towards investigations of transients on time scale of fractions of a second to a few seconds \citep{schaefer_1987, rotse_2002, karpov_2010, karpov_2019, richmond_2020, tingay_2020, arimatsu_2021}, they had only a limited success in detecting initial phases of gamma-ray bursts \citep{racusin_2008, karpov_grb_2017, zhang_2018}, in part due to quite limited depth of these surveys.

On the other hand, the zoo of rapid optical transients is potentially extremely diverse, spanning from short intense flares on red dwarf stars to gamma-ray bursts and fast radio bursts \citep{frb}.
And while modern and upcoming surveys like ZTF and LSST have insufficient temporal resolution and cadence to properly sample the light curves of such events, due to extreme sensitivity they still may detect both their peaks and afterglows, as it was demonstrated by the recent ZTF discovery of orphan GRBs \citep{ztf_orphans, ztfrest}.

The search for rapid optical transients on sub-second time scales is significantly complicated by the large background of artificial events -- satellite glints, or flares, caused by the sunlight reflection from solar panels or rotating antennae onboard artificial satellites orbiting the Earth. Such events may both have sub-second durations and be bright enough to be apparent even in smaller-scale wide-field surveys \citep{mmt_flashes, nir_2020, corbett_2020}. 
Even in deep surveys with lower temporal resolution, where satellites typically produce extended streaks and trails easily detectable by their morphology, such rapid flashes may still generate point sources
indistinguishable from stellar or extragalactic objects that may mimic e.g. gamma-ray bursts \citep{nir_2021}.

In this article we investigate the impact of such satellite glints 
on the transients detectable in the ZTF alert data stream using the data archived by the \fink\ broker \citep{10.1093/mnras/staa3602} between November 2019 and April 2023 (Section~\ref{sec:candidates}).
We use several criteria for associating the alerts with satellites, based either on purely geometric analysis of the transients detected on individual exposure, morphological properties of the alert cutouts, or results of direct association of transient positions with propagated ephemerides of known catalogued satellites (Section~\ref{sec:glints}). We then use these associations in order to derive the properties (duration and intrinsic brightness) of the flares, and estimate the all-sky rate of such events. Finally, we briefly discuss how these results may scale to LSST (Section~\ref{sec:discussion}).

\section{Selection of candidate events}
\label{sec:candidates}

For this study we used a subset of alerts from ZTF public alert stream \citep{Bellm_2018} which are being processed and archived by \fink\ community alert broker\footnote{\url{https://fink-broker.org}} \citep{10.1093/mnras/staa3602} since late 2019. This is an unfiltered, 5-sigma alert stream extracted from difference\footnote{Difference images are produced by ZTF pipeline \citep{Bellm_2018} by subtracting the reference image from every science image acquired during the survey. The reference, or template, images are produced by co-adding large number of highest quality science images, and are updated periodically as a part of ZTF data release process.} images that primarily includes events from flux transients, variables, and Solar System objects. Detectable streaks, such as aircraft and satellite trails, as well as cosmic rays, are removed by the ZTF processing pipeline \citep{Masci_2019} and \texttt{CreateTrackImage} software \citep{2014PASP..126..674L} prior to generating the alerts. Upon ingestion of alerts into \fink, two additional quality criteria (the real-bogus score\footnote{Real-bogus score is a machine learning based quality score for the reliability of a transient candidate based on a number of pixel-based features computed around transient position \citep{2019PASP..131c8002M}. It has a range of values between 0 and 1, where values closer to 1 imply a more reliable candidate.} being above 0.55, and there should be no prior-tagged bad pixels in a 5 x 5 pixel stamp around the alert position) are being applied.

In order to identify the events linked with artificial satellites, we started with a set of all non-repeating (i.e. appearing on just a single exposure, and not detected in consecutive observations on the same sky position) isolated alerts not connected to known stationary objects, as well as to known moving Solar System objects. To reject the latters, we excluded every alert that is within 30 arcseconds of current positions of objects from Minor Planet Center (MPC) database. 

\begin{figure}[t]
\centerline{\includegraphics[width=1.0\textwidth,clip=]{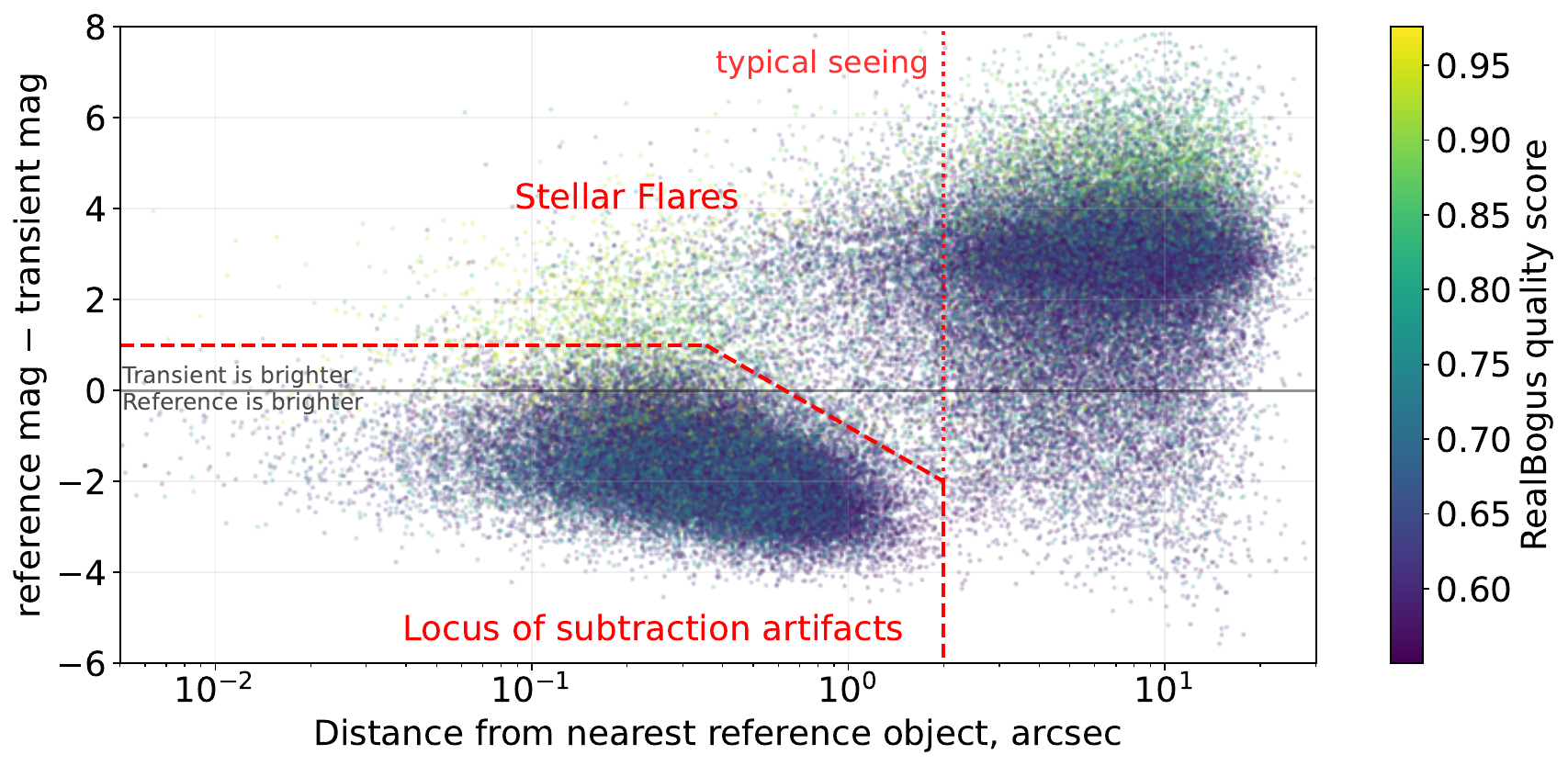}}
\caption{The difference between transient magnitude measured on difference image and the magnitude of nearest object from reference catalogue versus the distance to this nearest reference object for all events corresponding to non-repeating positive detections not coincident with Solar System objects. 
The value of RealBogus quality score 
is represented by the color of dots. Clearly visible cloud of lower-quality events in the lower part of the plot corresponds primarily to subtraction artefacts on stationary non-variable objects, as well as to occasional stellar variability like lower-amplitude stellar flares. Red dashed lines mark the extent of this ``stellar locus'' as used for rejection of these events.}
\label{fig:locus}
\end{figure}

To exclude the events related to stationary objects we located and excluded the ``stellar locus'' on the plane defined by the distance of the transient to the closest object in the reference catalogue and the difference of their magnitudes as shown in Fig.~\ref{fig:locus}. This locus contains the alerts related to both occasional stellar variability (i.e. small amplitude stellar flares) and -- most important for our analysis -- the image subtraction artefacts. The characteristic features of the latters are their innate proximity, within couple of typical PSF sizes, to the objects visible on reference frame (and thus populating the reference catalogue), and the brightness not exceeding the one of reference object. This locus contains on average 50\% percents of non-repeating alerts not connected to known Solar System objects.

This way we extracted a subset of 828063 candidate alerts, appearing on 164,432 of 252,532 (65.1\%) individual exposures acquired by ZTF over 815 observational nights between Nov 2019 and Apr 2023, with 53,581 of these exposures (21\% of all) containing 5 or more candidates.

These candidates constitute the natural sample for searching for both satellite glints and genuine rapid astrophysical transients as they contain all high-quality non-stellar and off-nuclear (i.e. located outside of centers of galaxies) alerts that manifest on just a single exposure, with most of the artefacts removed.

\section{Identification of satellite glints}
\label{sec:glints}

We employed several methods in order to isolate the candidates that are connected to flashes (glints) from artificial satellites among the candidates selected in Section~\ref{sec:candidates}. They are outlined in the following sub-sections.

\subsection{Detection of tracklets}
\label{sec:tracklets}

\begin{figure}[t]
\centerline{
\includegraphics[width=0.51\textwidth,clip=]{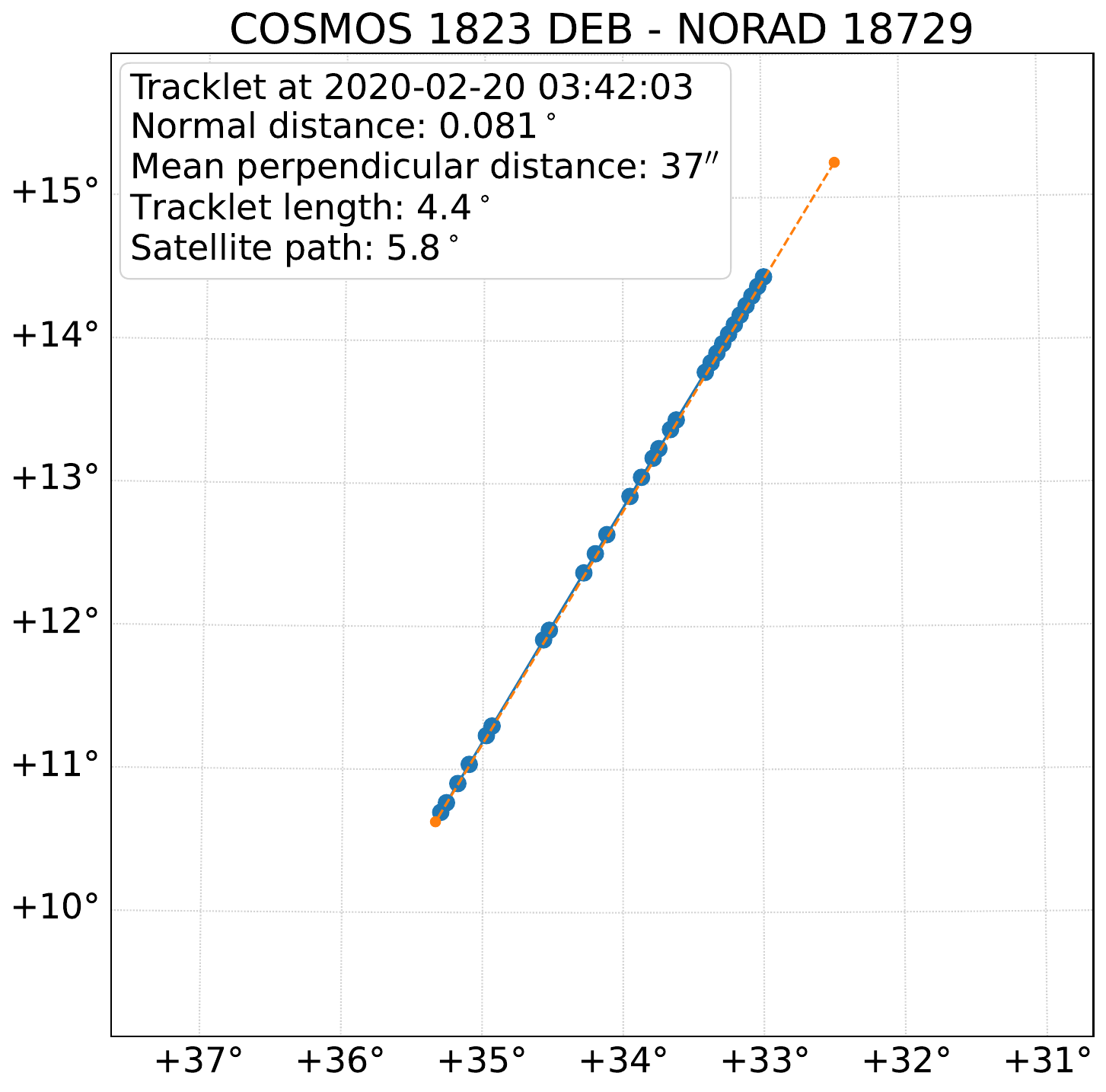}
\includegraphics[width=0.49\textwidth,clip=]{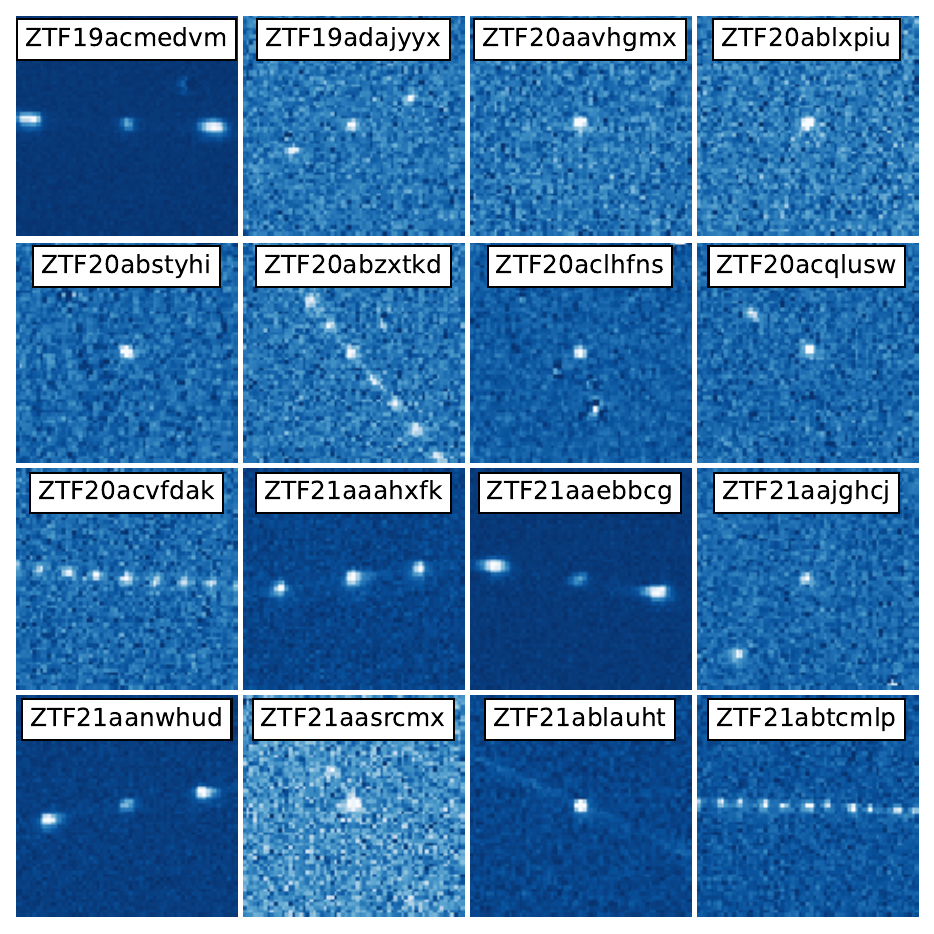}
}
\caption{Left panel -- example of a tracklet detected on a single ZTF exposure, plotted in celestial coordinates. The arc of a best matching satellite from NORAD catalogue during the exposure is also shown with orange dashed line. Right panel -- example cutouts of difference images from ZTF alert packets corresponding to a random subset of events associated with tracklets. Large part of them show characteristic multi-peak structure, often symmetric in respect to the ``main'' event, while some fraction is still essentially isolated point sources.}
\label{fig:methods}
\end{figure}

As significant fraction of all candidates appear on the frames in large groups, we attempted to detect quasi-linear structures among them -- to identify the sequences of events located approximately along the same great circle inside individual exposure. While a few points may occur along the line just randomly, for 5 or more events it is highly unlikely for typical numbers of events per exposure we have. Such sequences of events -- tracklets -- correspond to rapidly spinning satellites that produce numerous short flashes during the ZTF exposure, shifting between them due to orbital motion. 

The algorithm for detecting the tracklets starts with constructing great circles for every pair of candidates on the exposure, and keeping the ones with enough points close to them. The algorithm then gradually lifts the constraints on the distance and fits the second-order polynomial for the deviation from the great circle in order to properly capture longer tracks where linear approximation for orbital trajectory projection is no more sufficient. Finally, the algorithm merges together nearly co-linear tracks, and requires them to contain at least 5 points in order to be accepted.

Using this algorithm, we detected 10,170 individual tracklets that contain 116,389 (14.1\%) of all candidates.

\subsection{Cross-identification with NORAD satellite catalogue}
\label{sec:satellites}

We cross-matched detected tracklets with the positions of known satellites from public NORAD catalogue published by the Joint Space Operation Center (JSpOC). As the orbital elements rapidly evolve in time, we used sets of two-line elements (TLE) downloaded daily from  \url{https://www.space-track.org} for the analysis of tracklets detected on the same day. Then, we propagated these orbital elements to the times of individual exposures using {\sc Skyfield} package \citep{skyfield}. 
We then selected the satellites with arcs during the exposure approximately co-linear with the tracklet and not shorter than its length, and with normal distance of tracklet points from satellite arc not exceeding 300 arcseconds. For every tracklet we selected the satellite with smallest mean normal distance as a match.

This way, we associated 6,238 (61.3\%) of all tracklets with catalogued satellites. Example of such association is shown in left panel of Fig.~\ref{fig:methods}.

We also performed a ``blind'' association of individual candidates, regardless of them belonging to tracklets or not, with propagated positions of catalogued satellites in a similar manner, but now requiring the candidate to lay closer than 60$''$ from satellite arc, and to have the distance along the arc not exceeding its length, thus permitting for some acceptable error in the satellite fly-by time. This way, we acquired satellite associations for 95,279 (11.5\%) candidates. Among them, 71,032 (74.6\%) belong to the tracklets. In total, both variants (less restrictive one based on association of tracklets, and more strict one for matching positions of individual events) allowed us to associate 99,157 (12.0\%) candidates with known satellites from NORAD catalogue. 

\subsection{Morphology of alert images}
\label{sec:morphology}

Right panel of Fig.~\ref{fig:methods} shows how a random subset of events belonging to tracklets look like on cutouts from ZTF difference images, i.e. with all stationary objects (stars, galaxies, etc) removed. Significant fraction of them shows multiple appearances of the transient inside the same cutout (with ~60x60$''$ sizes) due to sufficiently rapid (quasi-) periodic variability due to fast rotation of the satellites.

In order to locate the events with such peculiar appearances in differential images 
we performed a simple morphological analysis of the cutouts\footnote{Cutout images distributed within ZTF alert packets contain small portions of both science, reference and difference images around transient positions.} of all candidates. To do so, we determined the regions of cutouts where brighter objects are apparent in the template images, and then performed peak detection above 2$\sigma$ level everywhere except these regions
in difference images using {\sc SEP} code \citep{sep}. 
This way, we successfully detected peaks for 114,051 (98.0\%) of all events belonging to tracklets, and for 86,790 (74.6\%) of them detected two or more peaks per cutout. Among all candidates, 151,046 (18.2\%) also displayed multiple peaks, which we may consider as a manifestation of satellite rotation.

\section{Discussion}
\label{sec:discussion}

\begin{figure}[t]
\centerline{
\includegraphics[width=0.48\textwidth,clip=]{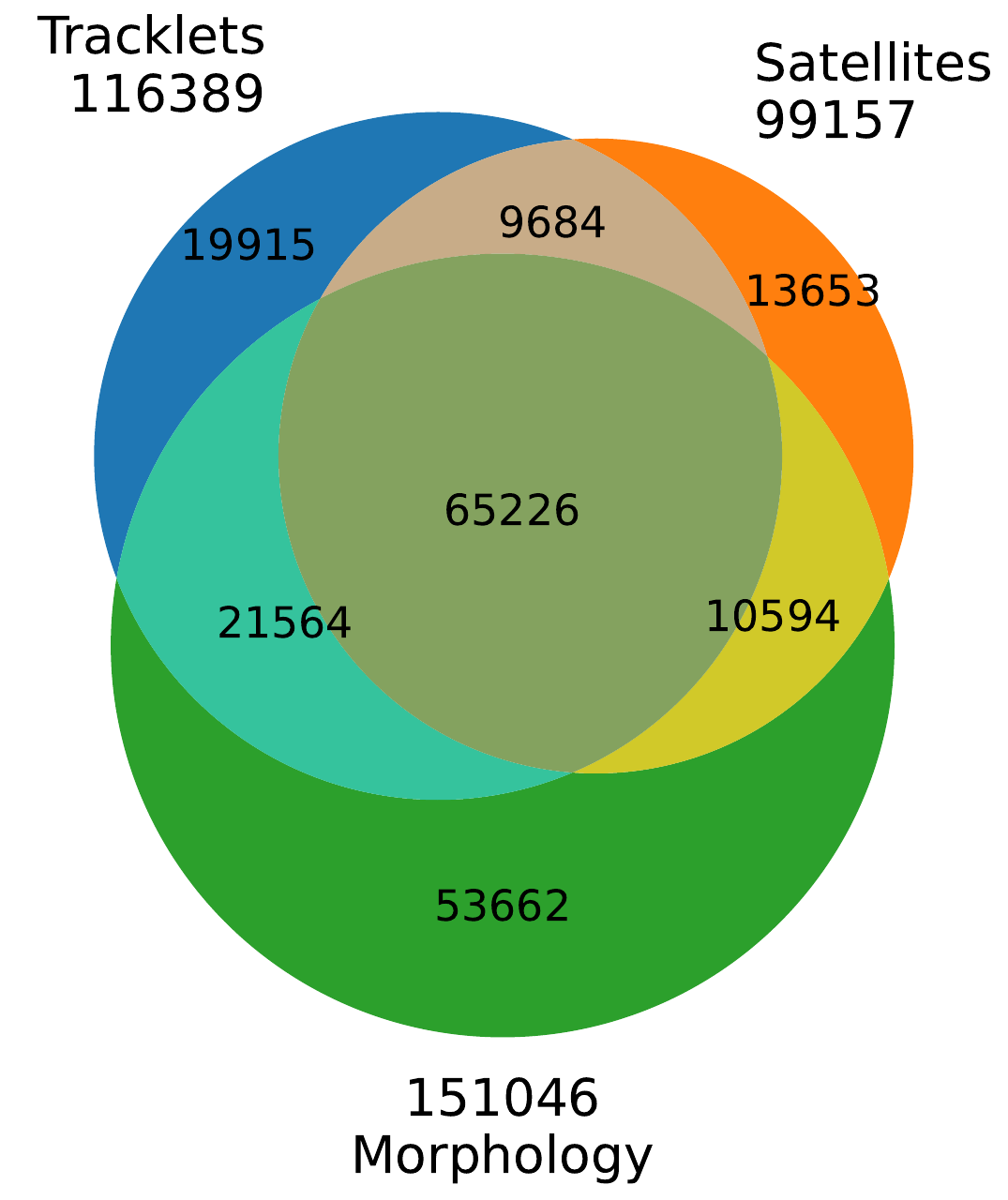}
\includegraphics[width=0.52\textwidth,clip=]{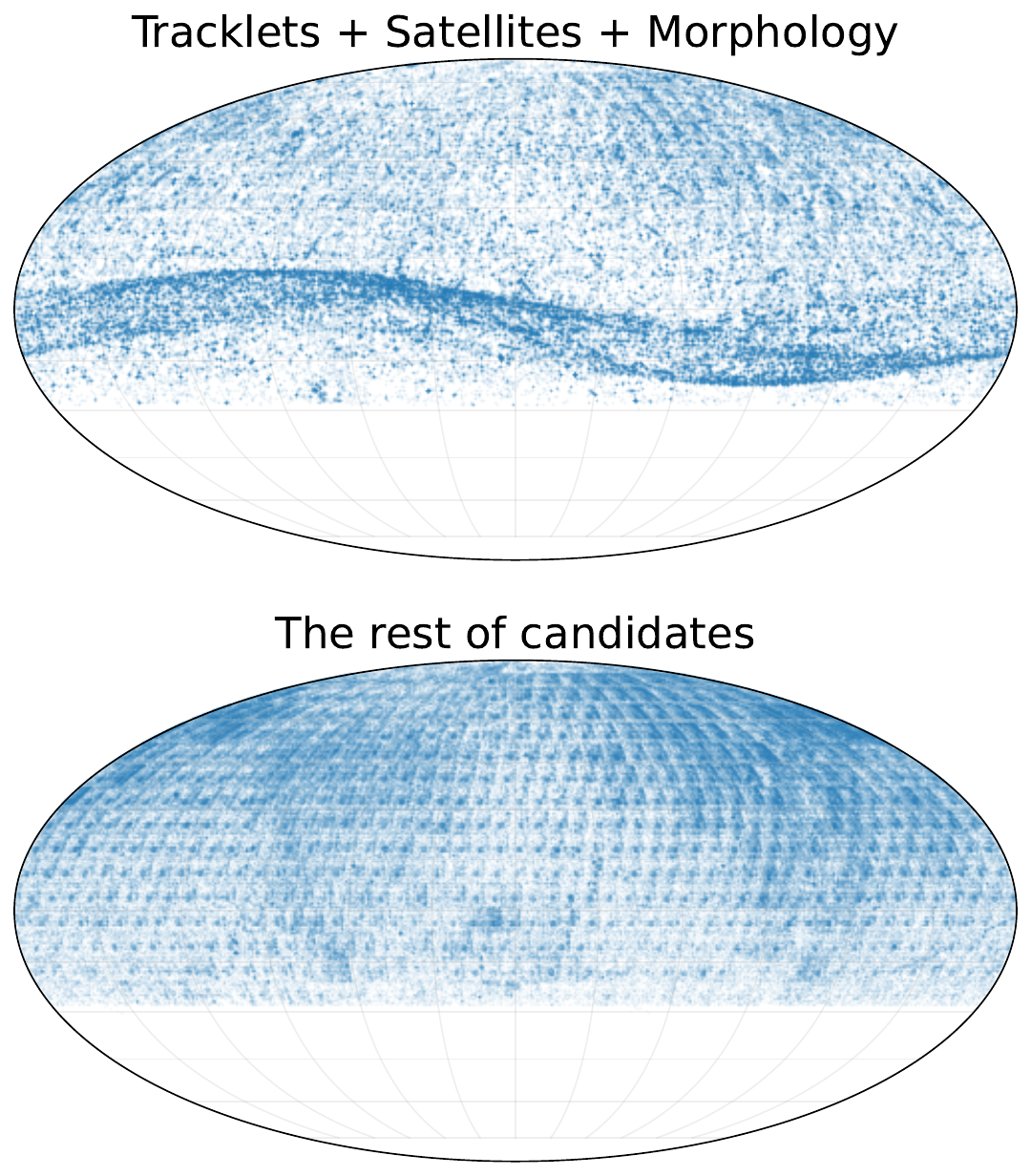}
}
\caption{Left panel -- Venn diagram of satellite associated events selected by three different methods as described in Section~\ref{sec:glints}. In total, these events constitute 23.5\% of all candidates. Right panel -- sky maps in equatorial coordinates showing the distributions of events both associated with satellites using one of three methods outlined in Section~\ref{sec:glints}, and not associated ones.}
\label{fig:venn}
\end{figure}

As described in Section~\ref{sec:glints}, we used three different methods for identifying the events corresponding to satellite glints. Each of them has its own shortcomings, and thus they select slightly different subsets of events. 
Left panel of Fig.~\ref{fig:venn} shows a Venn diagram of the relation between the sets selected by each of the methods. In total, 194,298 (23.5\%) candidates belong to either of three sets.
Their distribution over the sky (right panel of Fig.~\ref{fig:venn}) is significantly different from the rest of the candidates, and displays a prominent ``belt'' along the ecliptic equator, related to the flashes from geostationary satellites. The regular grid visible in the sky map of non-associated candidates is due to interplay of a fixed tiling strategy of ZTF survey observations, and significant excess of the events in several of 64 read-out channels of CCD mosaic covering ZTF focal plane. The latter suggest that it may be to some degree contaminated with noise events. The rest of these non-satellite associated events is probably genuinely astrophysical, and contains e.g. large amplitude flares from late-type stars too faint to be seen in reference images, but their analysis is outside the scope of current work.

\begin{figure}[t]
\centerline{
\includegraphics[width=0.99\textwidth,clip=]{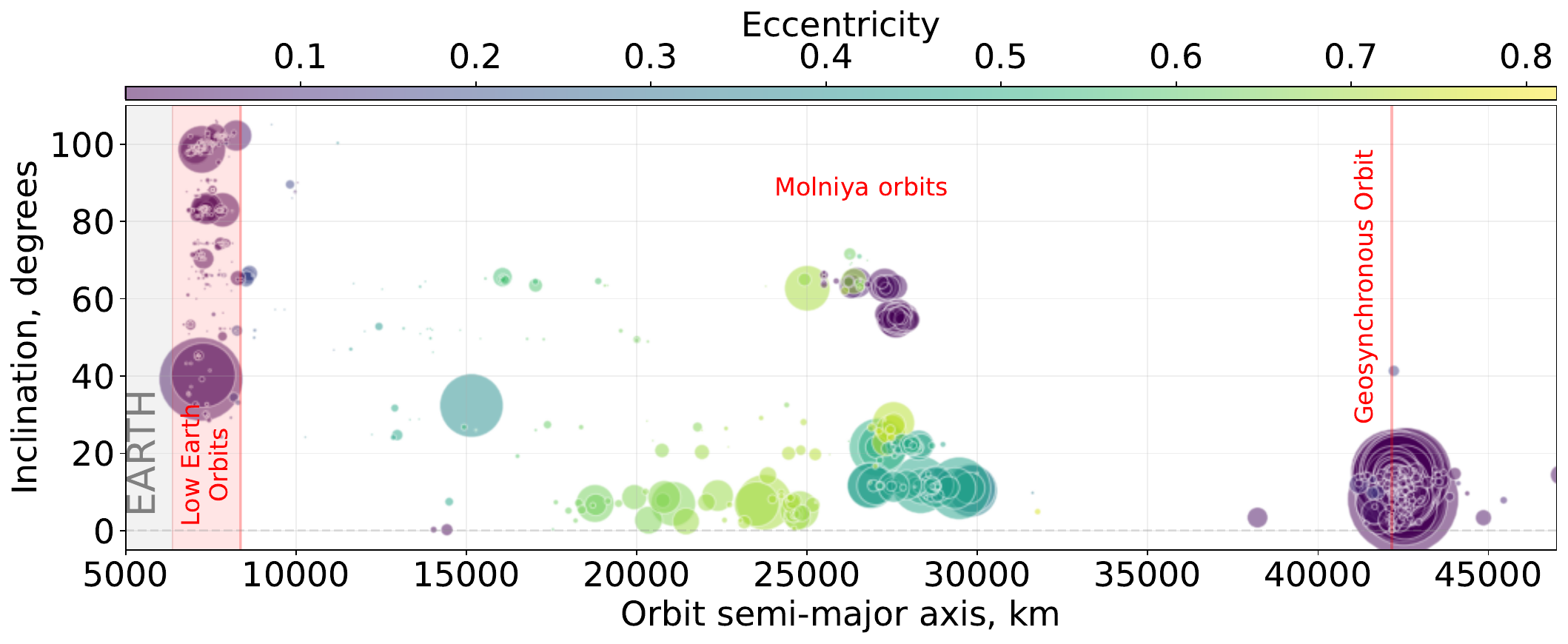}
}
\centerline{
\includegraphics[width=1\textwidth,clip=]{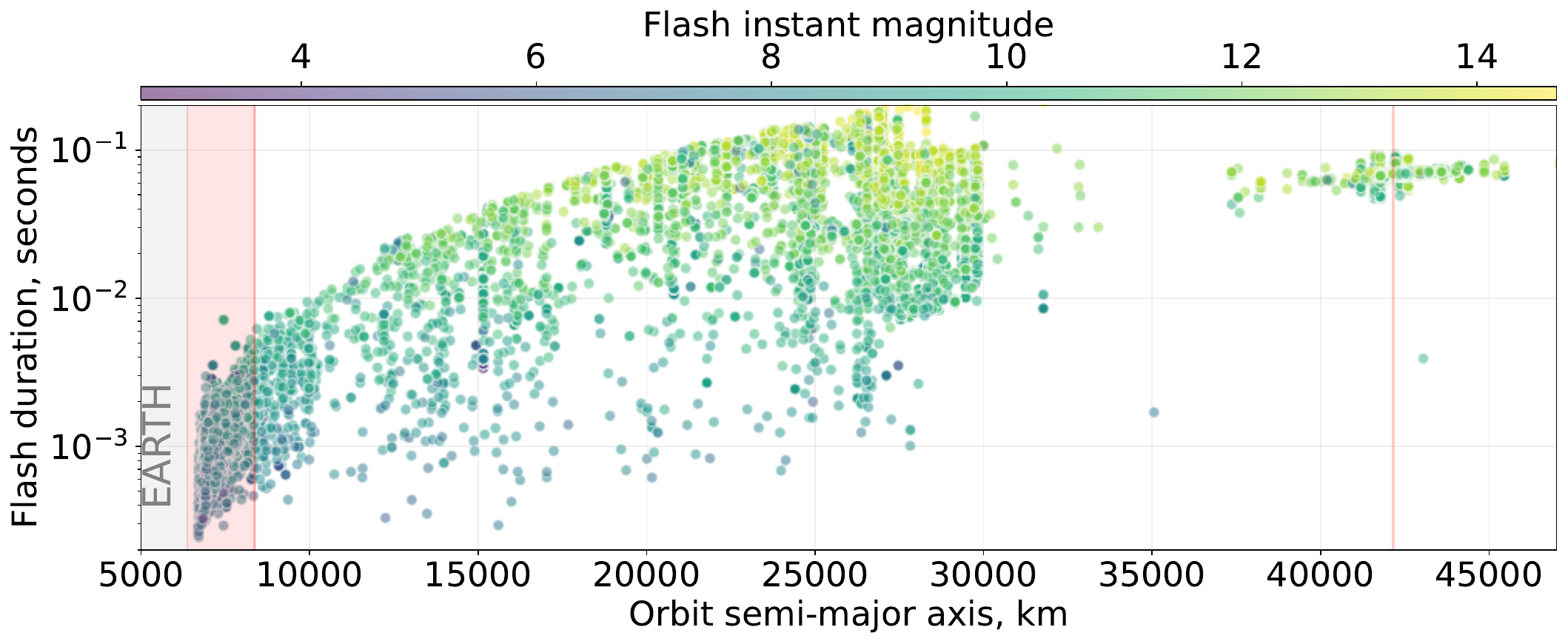}
}
\caption{Upper panel -- orbital parameters for the satellites from NORAD catalogue associated with tracklets as described in Section~\ref{sec:satellites}. Three zones of interest are displayed: Low-Earth orbits (with semi-major axis below an altitude of 2000 kilometers), \textit{Molniya} orbits (centered on a semi-major axis corresponding to half a sidereal day, 26,000 kilometers, and an inclination of 63.4 degrees), and geosynchronous orbits. Each circle corresponds to a satellite, with size proportional to the number of its appearances in the dataset.
Lower panel -- distribution of flash durations of the events associated with known satellites as a function of their orbit semi-major axis. 
Color-coded is the instant brightness of the flashes, as they would be measurable in observations with sufficiently high temporal resolution.
}
\label{fig:orbits}
\end{figure}

Upper panel of Fig.~\ref{fig:orbits} shows the distribution of orbital parameters for the catalogued satellites successfully associated with flashes in Section~\ref{sec:satellites}. They occupy all major loci in the parameter space -- near-Earth ones, geostationary satellites and highly inclined semi-synchronous \textit{Molniya}-type orbits, with some of the satellites observed multiple times and producing large numbers of flares. Most of these satellites are inactive and debris ones, where random rotation is expected. Small number of still active satellites in the sample is most probably being stabilized by rotation.

For these events associated with individual catalogued satellites we may also constrain the duration of the flashes by comparing the length of the satellite arcs during the ZTF 30-seconds exposure with typical size of a point source in the images. As the ZTF pixel size is 1$''$ (and typical seeing is 2$''$), we may derive the duration (actually, an upper limit for it) of the flash as
\begin{equation}
\tau = 30 \cdot \frac{1''}{\mbox{arc length}} \ \mbox{seconds.}
\label{eq:duration}
\end{equation}
On the other hand, instant brightness of the flash peak (the magnitude that would be measured if the temporal resolution of the observations would be high enough to fully resolve it) may be estimated as
\begin{equation}
\mbox{instant magnitude} = \mbox{mag} + 2.5\log{\frac{\tau}{30\ \mbox{seconds}}} \ \mbox{,}
\end{equation}
corresponding to the flashes being intrinsically brighter than they appear in ZTF images.
The distribution of the duration of the flashes, as well as their intrinsic brightness, estimated this way is shown in lower panel of Fig.~\ref{fig:orbits}. 

\begin{figure}[t]
\centerline{
\includegraphics[width=0.5\textwidth,clip=]{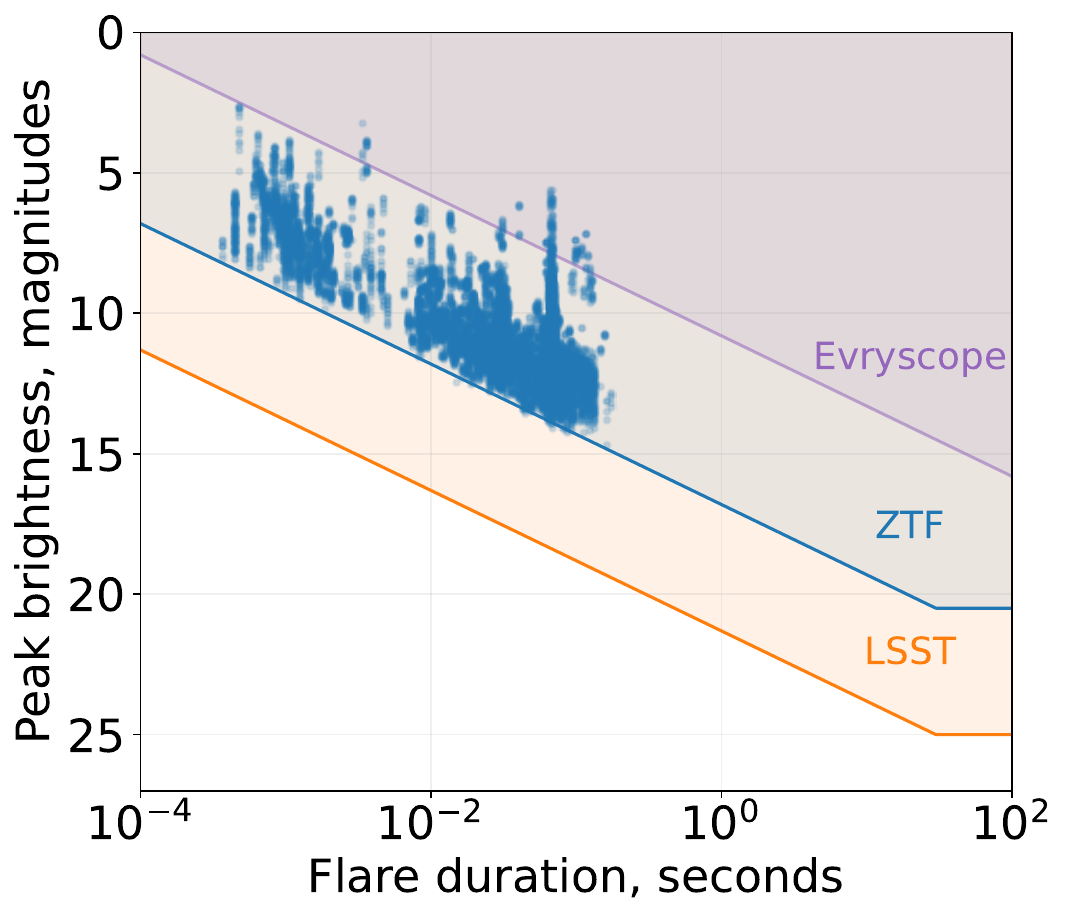}
\includegraphics[width=0.5\textwidth,clip=]{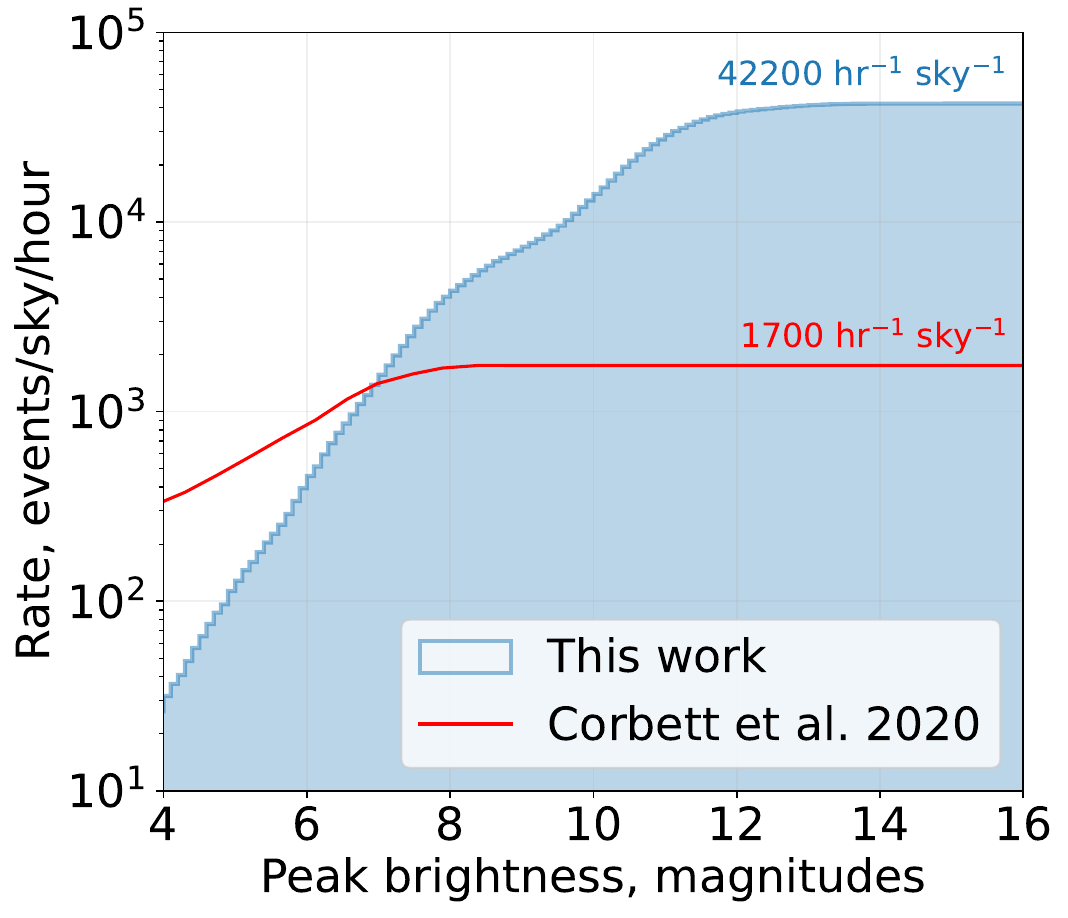}
}
\caption{Left panel --  distribution of flare duration and peak brightness for the events associated with catalogued satellites, together with detection limits for short flashes in three different sky surveys -- Evryscope having 120 s typical exposures \citep{corbett_2020}, ZTF \citep{Bellm_2018}, and upcoming Vera C. Rubin LSST \citep{lsst}, both having 30 s effective exposures. Right panel -- cumulative distribution of the flash rate averaged over the whole sky as derived in this work, in comparison with results of \citet{corbett_2020} based on Evryscope observations.}
\label{fig:rate}
\end{figure}

Effective detection limit of ZTF survey performed with a fixed exposure for the flashes of shorter duration is shown in left panel of Fig.~\ref{fig:rate}, along with the same for upcoming Vera C. Rubin LSST (which will have the same effective 30-s exposure as ZTF), and Evryscope \citep{corbett_2020}. Based on the data from the latter survey, \citet{corbett_2020} reported the average all-sky rate of the satellite flashes to be around 1,800 per hour for the peak brightness above magnitude 8. The values of peak brightness that we derived above for the events matched with catalogued satellites in Section~\ref{sec:satellites} may be used to update this rate estimate towards fainter flashes. The result is shown in right panel of Fig.~\ref{fig:rate}. The rate is up to 42,000 events per hour from the whole sky, with the majority of flashes having instant brightness between 10 and 12 mag. For the whole set of events caused by glinting satellites (see left panel of Fig.~\ref{fig:venn}) the all-sky rate is even higher, up to 80,000 events per hour.
The difference of the rates for brightest events between ZTF and Evryscope data is significantly affected by both different saturation limits of these surveys, and by the fact that intrinsically brightest flashes are predominantly produced by the low-orbit satellites, which are less likely to stay point-like in ZTF than in Evryscope (whose pixel size is 13$''$, so 13 times worse, while its typical 120 seconds exposure is only 4 times longer than ZTF one).

Extrapolating these results to upcoming LSST survey which is about to start in less than a year is not trivial, as its pixel scale and seeing will be significantly better (0.2$''$ and 0.8$''$) than the ones of ZTF (1$''$ and 2$''$), while the detection limit will be at least 4 magnitudes deeper.
The effect is twofold -- some flashes that are seen as point-like in ZTF may become elongated, while the effective pixel crossing time will be smaller, and thus the quiescent trail of the satellite between the flashes will become fainter. The distribution of upper limits for the flare duration in lower panel of Fig.~\ref{fig:orbits}, and recent study of fainter debris on geostationary orbits by \citet{debriswatch} that revealed a large number of uncatalogued objects with significant amount of ``tumbling'' ones, suggest that LSST will also be significantly affected by satellite glints that will impact the detection of rapid flares of astrophysical origin.

\section{Conclusion}
\label{sec:summary}

In order to constrain the rate of satellite glints as seen by modern and upcoming large-scale sky surveys we analyzed a subset of events ZTF alert stream that are detectable just on single exposure and are not caused by sporadic stellar variability or subtraction artefacts, thus potentially containing rapid optical transients of astrophysical origin. Using three different methodologies -- simple geometrical routine for associating the events on individual exposures into ``tracklets'' roughly along the the great circles, direct association of event positions with the arcs of catalogued NORAD satellites, and simple morphological analysis of the alert cutouts -- we associated 23.5\% of all these candidate events with satellite glints. 

The glinting satellites occupy all possible orbits around the Earth, from low to medium to geosynchronous ones, and the durations of the observed flashes are as short as $10^{-1}$--$10^{-3}$ seconds with instant brightness of $g$=4--14 magnitudes. The total rate of these flashes is on average up to 80,000 per hour from the whole sky, and is larger in the region along the celestial equator occupied by the flashes from geostationary satellites. Assuming that the proportion of objects orbiting the Earth increases with the decrease of their size, we expect deeper surveys such as the Vera C. Rubin LSST to detect much more events of this kind.


\acknowledgements
This work was developed within the Fink community and made use of the Fink community broker resources. Fink is supported by LSST-France and CNRS/IN2P3. This project has received financial support from the CNRS through the MITI interdisciplinary programs and from CNES.
SK acknowledges support from the European Structural and Investment Fund and the Czech Ministry of Education, Youth and Sports (Project CoGraDS-CZ.02.1.01/0.0/0.0/15003/0000437).


\bibliography{ref}

\end{document}